\documentclass[aps,pre,twocolumn,showpacs,floatfix,superscriptaddress]{revtex4-1}
\usepackage{graphicx}
\usepackage{amssymb}
\usepackage{dcolumn}
\usepackage{bm}
\usepackage{dcolumn}
\usepackage{amsmath}
\begin{document}
\title{Pore emptying transition during nucleation in hydrophobic nanopores}
\author{Milo\v{s} Kne\v{z}evi\'{c}}
%\email{milos.knezevic@physics.ox.ac.uk}
\author{Julia M. Yeomans}
\affiliation{The Rudolf Peierls Centre for Theoretical Physics, Oxford OX1 3NP, United Kingdom}
\date{\today}

\begin{abstract}
Using the 2D Ising model we study the generic properties of nucleation in hydrophobic nanopores. To explore the pathways to nucleation of a spin-up phase from a metastable spin-down phase we perform umbrella sampling and transition path sampling simulations. We find that for narrow pores the nucleation occurs on the surface outside the pore. For wide pores the nucleation starts in the pore, and continues outside the filled pore. Intriguingly, we observe a pore emptying transition for a range of intermediate pore widths: a pre-critical nucleus fills the pore, continues to expand outside of the filled pore, but then suddenly gets expelled from the pore before reaching its critical size.
\end{abstract}

%\pacs{83.30.Va, 61.41.+e, 61.30.-v, 78.20.H-, 81.40.Jj, 82.50.Hp}

\maketitle

\section{Introduction}

Efficient design of hydrophobic surfaces is one of the important contemporary challenges in fundamental and engineering sciences~\cite{Quere2008}. When a liquid droplet is placed on a hydrophobic, rough surface, it is typically found in one of two states: the droplet may be suspended on top of any pores trapping air underneath it, resulting in a Cassie-Baxter state~\cite{Cassie1944}, or wet the pores resulting in a Wenzel state~\cite{Wenzel1936}. The different fluid configurations result in different properties: in the superhydrophobic Cassie-Baxter state the drop has a higher contact angle and smaller hysteresis as the substrate-liquid contact is minimised. 

In a condensing environment, liquid condensate originates from nanometric nuclei and usually starts to grow within the pores thus giving rise to a Wenzel state which may alter the physical character of the rough surfaces. Indeed, dew-repellent surfaces found in nature~\cite{Barthlott1997,Chen2013} often exhibit a nanoscale roughness which obstructs formation of these nanometric nuclei in the cavities. The nanostructural dew repellency in living systems has inspired artificial surfaces that are both superhydrophobic and antidew~\cite{Chen2007,Miljkovic2012a}. Despite this progress, the antidew mechanism of the nanostructured surfaces remains unclear, hindering the applications of superhydrophobicity in a variety of systems such as effective condensers~\cite{Miljkovic2012b,Miljkovic2012c,Miljkovic2013,Narhe2004,Varanasi2009,Boreyko2009}, self-cleaning materials~\cite{Chen2013,Sun2005} and anti-icing substrates~\cite{Poulikakos2014}. The nanometric scale poses significant challenges in terms of experimental measurements, because it is very difficult to realise a sufficiently good temporal and spatial resolution to track the nucleation dynamics. To try to reveal the nucleation mechanism at these scales one instead resorts to theoretical predictions and computer simulations. To this end, some progress has already been made by Guo \textit{et al.}~\cite{Guo2014} who used a constrained lattice density functional theory to study the vapour-liquid nucleation of droplets on nanopillared structures in the lattice-gas model.

Our aim in this paper is to study the generic features of nucleation in hydrophobic pores. To this end we perform Monte Carlo simulations of the nearest-neighbour 2D Ising model to explore the pathways to nucleation of a spin-up phase from a metastable spin-down phase. The connection to the vapour-liquid transition can then be made by the standard mapping~\cite{Lee1952} of the Ising model to the lattice-gas model. This approach has proved useful in the study of homogeneous nucleation~\cite{Pan2004,Ryu2010r,Ryu2010}, heterogenous nucleation on impurities~\cite{Sear2006}, and heterogeneous nucleation in pores with neutral walls~\cite{Page2006} (which do not favour either the up or down spins).

We analyse the nucleation pathways for hydrophobic pores, that is pores with walls favouring spin-down states. Depending on the pore size, we find that there are three possible nucleation regimes.
For small pore widths, the nucleation takes place on the surface outside the pore (see Fig.~\ref{fgr:10}). For sufficiently large pore widths the nucleation occurs as a two-step process: a pre-critical nucleus forms in one of the bottom corners of the pore and there is a free energy barrier that must be overcome for the pore to be filled; then there is a second nucleation outside the filled pore characterised by a critical nucleus which forms on top of the filled pore (Fig.~\ref{fgr:14}). This two-step nucleation regime has already been observed~\cite{Page2006} for pores with walls which are neutral. Finally, for moderate pore widths, we observe a pre-critical nucleus that starts to grow in the pore, fills the pore and proceeds to grow out of the pore without reaching a critical size. Then at a given moment it is suddenly expelled from the pore (see Fig.~\ref{fgr:12}) and, after some further growth, forms a critical nucleus that lies on the surface outside the pore as in Fig.~\ref{fgr:10}. Thus, in this nucleation regime one observes a pore emptying transition of the pre-critical nucleus.

The article is organised as follows. We first present the model in Section \ref{model}. Nucleation on a flat hydrophobic surface is studied in Section \ref{flat} using umbrella sampling and transition path sampling. In Section \ref{neutral} we briefly examine the case of neutral pores using umbrella sampling to compare with the results of Ref. [\onlinecite{Page2006}]. We present our results on nucleation from hydrophobic pores in Section \ref{hydrophobic}. A summary of our findings is given in the final section. 

\section{The model}\label{model}

\begin{figure}[h]
  \centering
  \includegraphics[width=8.6cm]{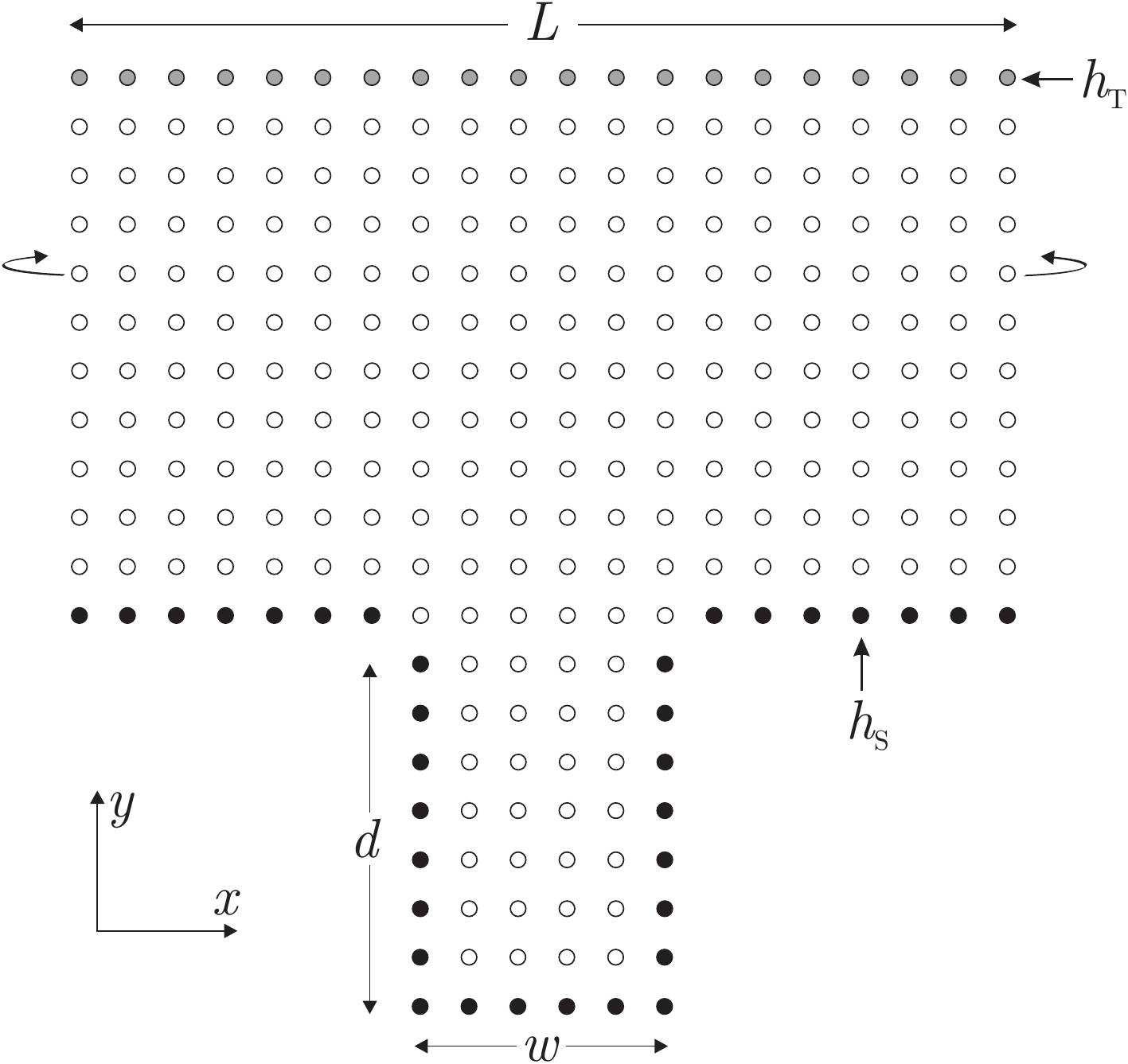}
  \caption{A schematic of the pore geometry, with $L$ the lattice size, $d$ the pore depth and $w$ the pore width (distance is measured in number of lattice vertices). Periodic boundary conditions are used in the $x$ direction. The surface field $h_{\textnormal{S}}$ acts on spins marked with dark grey, while surface field $h_{\textnormal{T}}$ acts on spins marked with light grey.}
\label{fgr:1}
\end{figure}

We study nucleation in the simple pore geometry shown in Fig.~\ref{fgr:1}. The lattice has $N = L \times L$ spins with periodic boundary conditions in the $x$ direction and closed boundaries in the $y$ direction.
Portions of the lattice are removed to give pores of depth $d$ and width $w$. Each lattice site $i$ is occupied by an Ising spin $s_i = \pm 1$.
The system is described by the Hamiltonian
\begin{equation}
	\mathcal{H} = - J \sum_{\langle i,j \rangle} s_{i} s_{j} - h \sum_{i} s_{i} - h_{\textnormal{S}} \sum_{i \in \: \textnormal{PS}} s_{i} - h_{\textnormal{T}} \sum_{i \in \: \textnormal{TR}} s_{i},
\end{equation}
where $J>0$ is the exchange coupling constant, $h$ is the bulk magnetic field, and $h_{\textnormal{S}}$ and $h_{\textnormal{T}}$ are additional surface magnetic fields.
The first term describes interactions between spins, and the bracketed sum over $i$ and $j$ indicates a restriction to nearest-neighbour pairs. The second term represents the coupling of all spins $s_{i}$ with the uniform external bulk magnetic field $h$.
The third term is the coupling of spins located at the pore surface (PS; dark grey in Fig.~\ref{fgr:1}) with the surface magnetic field $h_{\textnormal{S}}$, while the last term describes the coupling of spins in the top row (TR; light grey in Fig.~\ref{fgr:1}) of the lattice with an additional surface magnetic field $h_{\textnormal{T}}$. All spins in PS and TR have three nearest neighbours, except for the two spins located in the bottom corners of the pore which have only two neighbours.  

We measure the temperature in units of $J/k_{\textnormal{B}}$ and magnetic fields in units of $J$; here $k_{\textnormal{B}}$ is the Boltzmann constant. At $h=0$ the 2D Ising model undergoes a continuous transition from the disordered phase to one of the two ordered phases~\cite{Onsager1944} at a critical temperature $T_{c} = 2.27 J/k_{\textnormal{B}}$. We set the temperature to be $T= 1.25 J/k_{\textnormal{B}}$, well below $T_{c}$.
We start in the spin-down phase but choose a positive bulk magnetic field $h= J/16$, so that the initial configuration is metastable. The system evolves following a Metropolis single spin-flip Monte Carlo (MC) algorithm with random choice of trial spin. Time is measured in units of MC sweeps. For our choice of $T$ and $h$ the nucleation of the spin-up phase from the spin-down phase is typically not rapid, and brute-force simulations of the system are prohibitively slow. We thus apply rare event sampling methods, umbrella sampling and transition path sampling (TPS), to gain insight into nucleation of the stable phase. 

We set $L=60$ as this turns out to be sufficiently large to avoid artefacts from finite lattice size for our choice of $T$ and $h$. Tuning the surface field $h_{S}$ allows us to set the contact angle $\theta$ of the nucleating spin-up cluster on a flat surface: one obtains a hydrophilic surface ($\theta < 90^{\circ}$) for $h_{S}>0$, and a hydrophobic surface ($\theta > 90^{\circ}$) for $h_{S}<0$. The case $h_{S} = 0$, $\theta = 90^{\circ}$, which does not favour either spin-up or spin-down, was previously examined in Ref. [\onlinecite{Page2006}]. The additional surface field $h_{T}$ in the top row of the lattice is set to a high negative value, thus ensuring that the nucleation always occurs at the bottom surface. 

\section{Nucleation on a flat surface}\label{flat}

\subsection{Classical nucleation theory}

According to the classical theory of homogeneous nucleation~\cite{Debenedetti1996,Oxtoby1992,Sear2007} for a 2D system the free energy needed to form a nucleus of $n$ molecules of the stable phase can be written
\begin{equation}
	\Delta G = - n \lvert \Delta \mu \rvert + \gamma C,
\end{equation}
where $\Delta \mu <0$ is the difference between the chemical potentials of the stable and metastable phases under bulk conditions, $\gamma$ is the macroscopic surface tension between the two phases, and $C$ is the circumference of the growing nucleus.
For a circular nucleus of radius $r$, $n = \rho_{n} \pi r^2$, and $C= 2r \pi$, where $\rho_{n}$ is the number density of the nucleating phase. For the Ising model the chemical potential difference is~\cite{Ryu2010r} $\vert \Delta \mu \rvert \approx 2h$, which gives
\begin{equation}
	\Delta G = - 2h n + 2 \left ( \frac{\pi n}{\rho_n}  \right )^{1/2} \gamma.
\label{CNT}
\end{equation}
The surface tension $\gamma$ is given as a function of temperature $T$ by Onsager's exact expression~\cite{Onsager1944}. The expression (\ref{CNT}) holds for a noncircular nucleus as well, provided that the surface tension $\gamma$ is substituted with an ``effective'' surface tension~\cite{Shneidman1999}.
The free energy of the nucleating cluster depends on the reaction coordinate, the nucleus size $n$. In our study $n$ represents the size of the largest cluster consisting of $+1$ spins. We consider a set of $+1$ spins to form a cluster if each spin in the set is a nearest neighbour of at least one other spin.

The critical nucleus $n^*$ corresponds to the maximum of the free energy barrier $\Delta G^*$ separating the metastable and stable states. From (\ref{CNT}) one finds
\begin{equation}
	n^* = \pi \left ( \frac{\gamma}{2h \rho_{n}^{1/2}} \right)^2, \quad \Delta G^* = \frac{\pi \gamma^2}{2h \rho_{n}}.
\end{equation}
Nuclei containing fewer than $n^*$ particles shrink spontaneously due to their large circumference to surface ratios, while nuclei larger than $n^*$ tend to spontaneously grow as the bulk free energy contribution dominates. 

Extension of the classical theory to the case of heterogeneous nucleation predicts that the interaction of the nucleus with the surface dramatically reduces the free energy barrier for nucleation~\cite{Debenedetti1996}, resulting in heterogeneous nucleation being dominant over homogeneous nucleation. For the simplest case of a uniform infinite planar surface the classical theory~\cite{Turnbull1950} envisages that the barrier to heterogeneous nucleation
$\Delta G_{\textnormal{surface}}^{*}$ differs from the homogeneous one by a function solely of the contact angle, $f(\theta)$:
\begin{equation}
	\Delta G_{\textnormal{surface}}^{*} = \Delta G^{*} f(\theta),
\label{hetnuc}
\end{equation}
where $0 \le f(\theta) \le 1$, with limits $f(\theta = 0^{\circ}) = 0$ and $f(\theta = 180^{\circ}) =1$. In the case of a neutral surface the theory predicts $f(\theta = 90^{\circ}) = 1/2$. 

\subsection{Umbrella sampling}
\label{umbrella}

We determine the free energy barrier for nucleation using umbrella sampling~\cite{Torrie1977,Chandler1987}. We perform the Ising model Monte Carlo simulations in the $(N, h, T)$ ensemble, which corresponds to the $(V, \mu, T)$ ensemble in the lattice gas model. The probability for the formation of sufficiently large clusters (typically $n>20$ in our simulations) can be approximated by the probability to find one such cluster in the system~\cite{tenWolde1998,Auer2004}, as the simultaneous occurrence of two large clusters is very unlikely. Thus we choose umbrella sampling windows with bias potentials that constrain the size of the largest cluster in a given system configuration. 

To do this a system configuration $x = \{ s_i \}_{i=1, 2, \dots}$ has, in each umbrella window $k$, a potential energy $U_k (x)$ which differs from the 
Hamiltonian by a weighting umbrella bias potential $W_k (x)$,
\begin{equation}
	U_k (x) = \mathcal{H}(x) + W_k (x).
\end{equation}
This bias potential is chosen to be a harmonic function of the size of the largest $+1$ spin cluster $n(x)$ in the given configuration,
\begin{equation}
	W_k [n(x)] = \frac{1}{2} \kappa [ n(x) - n_k]^2.
\end{equation}
The harmonic constant $\kappa$ is set to the same value, $\kappa = 0.2$, in all windows, and the minimum $n_k$ determines which cluster sizes are sampled preferentially in each umbrella window $k$. Thus the Monte Carlo simulation is split into a number of smaller simulations with overlapping umbrella windows. 

To implement the umbrella sampling we start the system in the first umbrella potential centred about $n_1 = 0$. After equilibrating in the potential $U_1(x)$ for $5\times 10^4$ time steps, we sample the equilibrated system for $5\times 10^5$ time steps and record the size of the largest cluster $n$ at each step. We follow Ref. [\onlinecite{Auer2004}] by performing Metropolis MC moves according to the Hamiltonian $\mathcal{H}(x)$, and then additionally deciding whether to accept or reject all these moves according to the pure biasing potential $W_k(x)$ after every 5 MC sweeps. The minimum of the bias potential is then incremented and the whole procedure is repeated for the next
umbrella window. The chosen sequence of bias potential minima is $n_k = 0, 5, 10 \dots$ up to sizes significantly greater than the critical cluster size.

We use the multi-state Bennet acceptance ratio method of Shirts and Chodera~\cite{Chodera2008} to estimate the free energies $G_k (T,h)$ corresponding to the potential $U_k$ in the umbrella window $k$. The method allows one to calculate the ensemble average of any observable $A(x)$ with respect to the Hamiltonian $\mathcal{H}(x)$. We divide the space of cluster sizes $n$ into bins. The probability
for the formation of a cluster of size $n$ falling into $i$-th bin is given by the expectation with respect to $\mathcal{H}(x)$,
\begin{equation}
	p_i= \langle \chi_i [n(x)] \rangle ,
\end{equation} 
where $\chi_i [n(x)]$ is an indicator function that takes the value of 1 if the size of the largest cluster $n(x)$ in configuration $x$ falls into bin $i$, and is zero otherwise. The free energy $\Delta G(n)$ for the formation of a cluster of size $n$ can then be computed (up to an irrelevant additive constant) from the probability $p(n)$ that the largest cluster has size $n$:
\begin{equation}
	\Delta G(n) = - k_{\textnormal{B}} T \ln [p(n)].
\end{equation}
We set the unknown additive constant by requiring that $\Delta G = 0$ in the first bin.
\begin{figure}[h]
\centering
  \includegraphics[width=8.6cm]{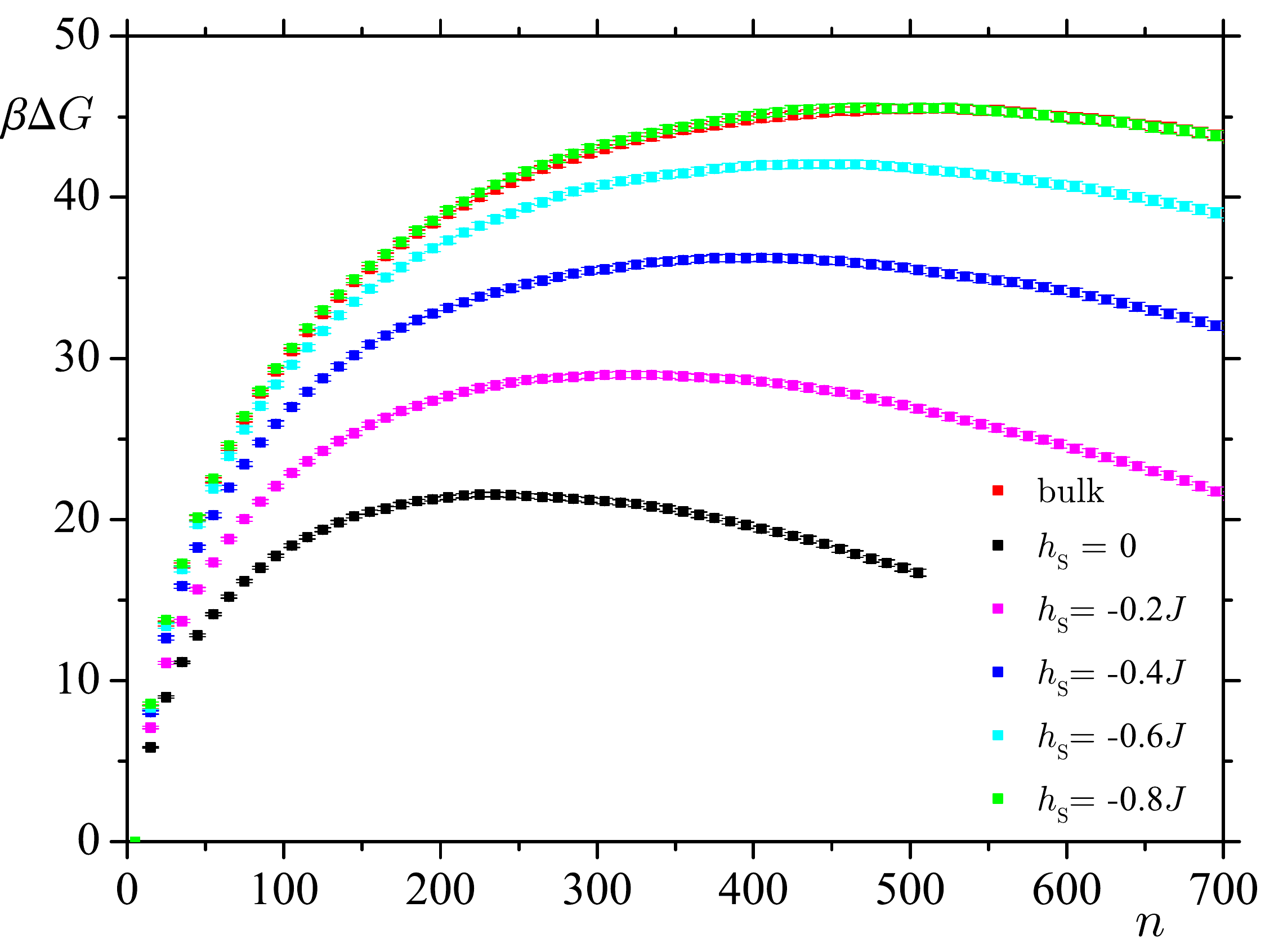}
  \caption{Free energy barriers for homogeneous and heterogeneous nucleation for $T=1.25J/k_{\textnormal{B}}$ and $h=J/16$. Free energy curves for $h_{\textnormal{S}} \le -0.8J$ practically coincide with the homogeneous free energy curve (red).}
  \label{fgr:2}
\end{figure}

Our umbrella sampling results for homogeneous nucleation and for heterogeneous nucleation on a flat surface are shown in Fig.~\ref{fgr:2}. In all simulations $T=1.25J/k_{\textnormal{B}}$ and $h=J/16$. The barrier for homogeneous nucleation is shown in red. Its height is $\beta \Delta G_{\textnormal{bulk}}^{*} = 45.52 \pm 0.27$ (here $\beta = 1/ k_{\textnormal{B}}T$), while the size of the critical nucleus is $n_{\textnormal{bulk}}^{*} \approx 515$. Here we do not attempt to fit the homogeneous nucleation free energy curve obtained in simulations to the classical theory. This is because it was shown in Ref. [\onlinecite{Ryu2010r}] that, for the 2D Ising model, the expression (\ref{CNT}) should be corrected to include an additional term which accounts for the shape fluctuations of the clusters to obtain a satisfactory match between the simulations and theory. 

We varied the surface field $h_{\textnormal{S}}$  from the value $h_{\textnormal{S}} = 0$ to a negative value $h_{\textnormal{S}} = -J$. The results for a neutral surface $(h_{\textnormal{S}} = 0)$ are shown in black. The barrier height is $\beta \Delta G_{\textnormal{neutral}}^{*} = 21.55 \pm 0.14$ and the size of the critical nucleus is $n_{\textnormal{neutral}}^{*} \approx 235$.
This case corresponds~\cite{Winter2009} to a contact angle of $90^{\circ}$ and, according to (\ref{hetnuc}), $\Delta G_{\textnormal{neutral}}^{*} = \Delta G_{\textnormal{bulk}}^{*}/2$.
In our simulations $\Delta G_{\textnormal{bulk}}^{*}/ \Delta G_{\textnormal{neutral}}^{*} = 2.1$, which is in surprisingly good agreement with (\ref{hetnuc}), given its simple nature. As the surface field $h_{\textnormal{S}}$ is decreased, the barriers for nucleation become higher, until they eventually reach the barrier for homogeneous nucleation for $h_{\textnormal{S}} \approx -0.8J$ where the energies for flipping a spin at the surface and in the bulk become equal. Further decrease of the field $h_{\textnormal{S}}$ does not alter the nucleation barrier, as the nucleation no longer occurs on the surface, but rather occurs in the bulk. In the opposite case, when the surface field is increased above the value $h_{\textnormal{S}} = 0$, the nucleation barriers become lower than the barrier shown in black in Fig.~\ref{fgr:2}, and eventually become negligible (not shown in the figure). 

\subsection{Transition path sampling}

We use transition path sampling (TPS)~\cite{Dellago2001,Bolhuis2002,Bolhuis2010} to determine the contact angles $\theta$ corresponding to the surface fields shown in Fig.~\ref{fgr:2}. TPS also provides an independent check on our estimates for the sizes of the critical nuclei. This method utilises a Monte Carlo walk in the space of reactive trajectories, connecting the metastable spin-down state and the nucleating spin-up state, to sample multiple examples of rare nucleation events. 
We use the size of the largest $+1$ spin cluster $n(x)$ in the system configuration $x$ as the order parameter, which discriminates between the initial and product states. Further, we define the TPS characteristic functions~\cite{Dellago2001,Bolhuis2002,Bolhuis2010} $h_{\downarrow}(x)$ and $h_{\uparrow}(x)$, which indicate whether a configuration $x$ is in the spin-down and spin-up state, respectively:
\begin{equation}
	h_{\downarrow}(x) = \begin{cases} 
      1, & n(x) < n_{\downarrow} \\
      0, & n(x) \ge n_{\downarrow} \\
   \end{cases}, \quad
         h_{\uparrow}(x) = \begin{cases} 
      0, & n(x) \le n_{\uparrow} \\
      1, & n(x) > n_{\uparrow} \\
   \end{cases},
   \label{regions}
\end{equation}
where $n_{\downarrow}$ and $n_{\uparrow}$ are the largest cluster size limits chosen to define the spin-down and spin-up states. In our TPS simulations we choose $n_{\downarrow}=20$ and $n_{\uparrow} = 600$ or $700$ (see Table \ref{tbl:1}) such that the initial state and product state basins do not overlap, and are both far enough from the transition state region found in the umbrella sampling simulations described in Section \ref{umbrella} (Fig.~\ref{fgr:2}). \\

With our choice of $T$ and $h$ we are in a regime where a nucleation event is very rare, making it extremely difficult to sample even one such event with brute force simulations. Thus we constructed the initial trajectory for TPS by taking an umbrella sampling configuration with $n = n^{*}$ and generating trajectories, with both forward and backward propagation using the underlying dynamics of the system, until a trajectory was found connecting the initial and nucleating states. 

Trajectories of 900 time units in length are sampled using the TPS shooting method~\cite{Bolhuis2010}. This trajectory length turned out to be long enough to harvest a sufficiently large number of transition states (critical clusters).

The TPS shooting move consists of choosing a random configuration $x_{\textnormal{r}}$ on the current  trajectory and choosing randomly to evolve the system either forwards or backwards. The new trajectory is then constructed by replacing the relevant part of the current path by the newly generated configurations. The updated trajectory is accepted if it connects the spin-down and spin-up states specified by functions (\ref{regions}).
The initial trajectory, constructed with the help of umbrella sampling data, may be unphysical, so we first equilibrate it with 2500 shooting moves, and then harvest about 60 independent trajectories, one every 50 accepted shooting moves. 
\begin{figure}[h]
\centering
  \includegraphics[width=8.6cm]{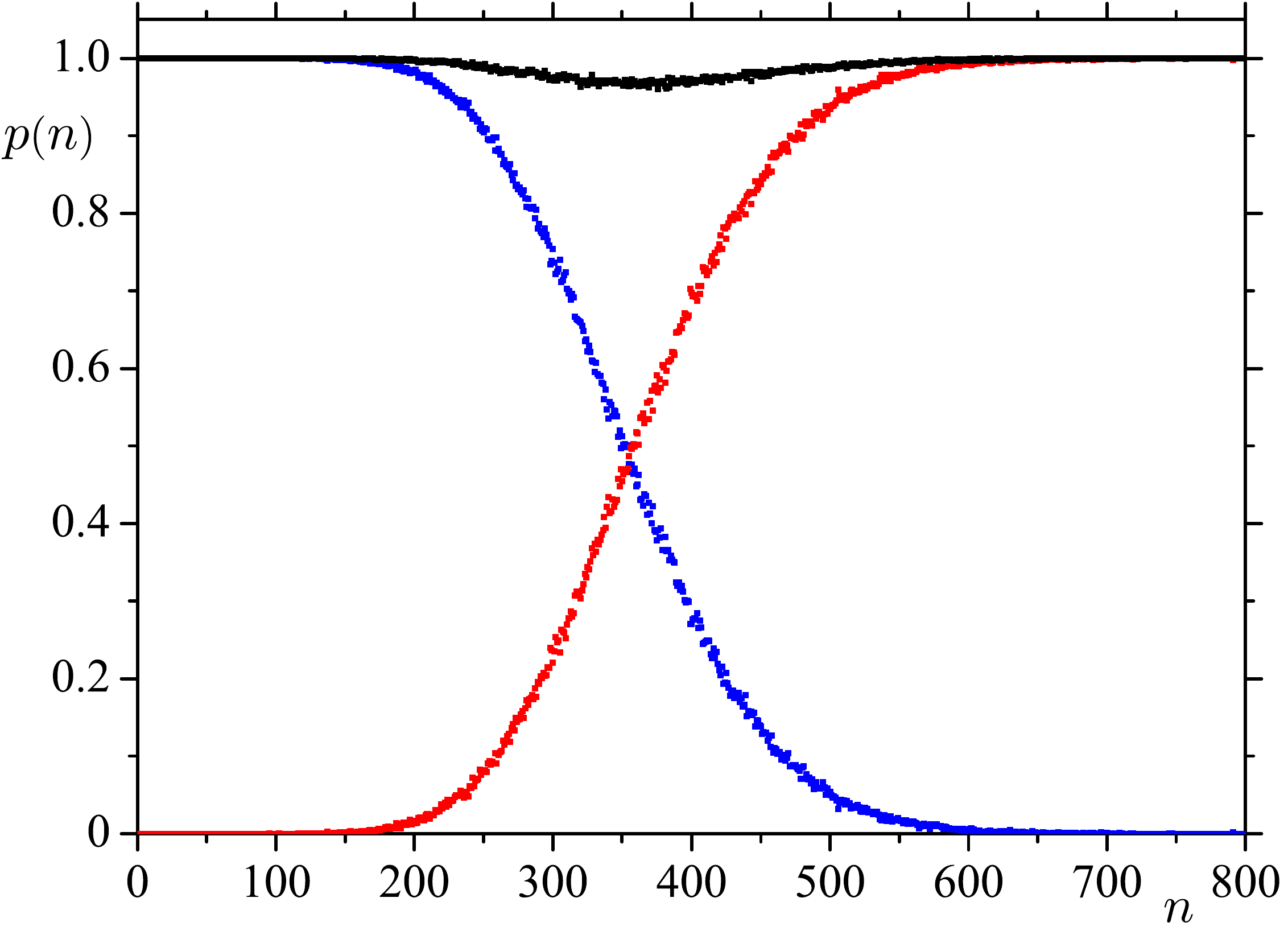}
  \caption{Committors for the case $h_{\textnormal{S}}= -0.3$: $p_{\downarrow}(n)$ -- blue line; $p_{\uparrow}(n)$ -- red line; and their sum -- black line. Fleeting trajectory lengths are $t_{\textnormal{s}} = 1800$ MC sweeps.}
  \label{fgr:3}
\end{figure}

We use the sampled ensemble of accepted trajectories to calculate the committors~\cite{Dellago2001,Bolhuis2002,Bolhuis2010} $p_{\downarrow}(n,t_{\textnormal{s}})$ and $p_{\uparrow}(n,t_{\textnormal{s}})$ for the spin-down and spin-up states, respectively. The committor $p_{\uparrow (\downarrow)}(n,t_{\textnormal{s}})$ is the probability that a system with initial configuration with a nucleus of size $n$ will reside in the spin-up (spin-down) state at time $t_{\textnormal{s}}$.
From each point of the sampled accepted trajectories we run 50 fleeting trajectories of length $t_{\textnormal{s}} = 1800$ time units. The spin-down and spin-up committors are then determined as the fraction of fleeting trajectories that end in the spin-down and spin-up regions respectively for each $n$. Our results for $h_{\textnormal{S}} = -0.3$ are shown in Fig.~\ref{fgr:3}. The committor graphs are qualitatively similar for all values of $h_{\textnormal{S}}$.
In general, not all fleeting trajectories end in spin-down or spin-up states; some of them might end in the region between these states. However, the fact that $p_{\downarrow}(n,t_{\textnormal{s}}) + p_{\uparrow}(n,t_{\textnormal{s}}) \approx 1$ for all $n$ in our simulations shows that the chosen fleeting trajectory length $t_{\textnormal{s}}$ is long enough to correctly sample the transition states. The transition region is defined as that consisting of states $n^{*}$ that are equally accessible, meaning $p_{\downarrow}(n^{*}) \approx p_{\uparrow}(n^{*})$. The results we obtain for $n^{*}$ are in good agreement with the corresponding umbrella sampling $n^{*}$ values for all surface fields $h_{\textnormal{S}}$.

A more general procedure would be to determine committors $p_{\downarrow}(x,t_{\textnormal{s}})$ and $p_{\uparrow}(x,t_{\textnormal{s}})$ for all system configurations $x$. However this would present an extraordinary computational challenge. Fortunately, for our needs it is enough to examine only the configurations $x$ of the ensemble of accepted trajectories connecting the spin-down and spin-up states. 
We follow the procedure described in Ref. [\onlinecite{Dellago2001}] to determine members of the transition state ensemble (TSE). TSE consists of states $x$ such that half of the fleeting trajectories initiated from $x$ end in spin-down state, and the other half end in spin-up state.
\begin{figure}[h]
\centering
  \includegraphics[height=5cm]{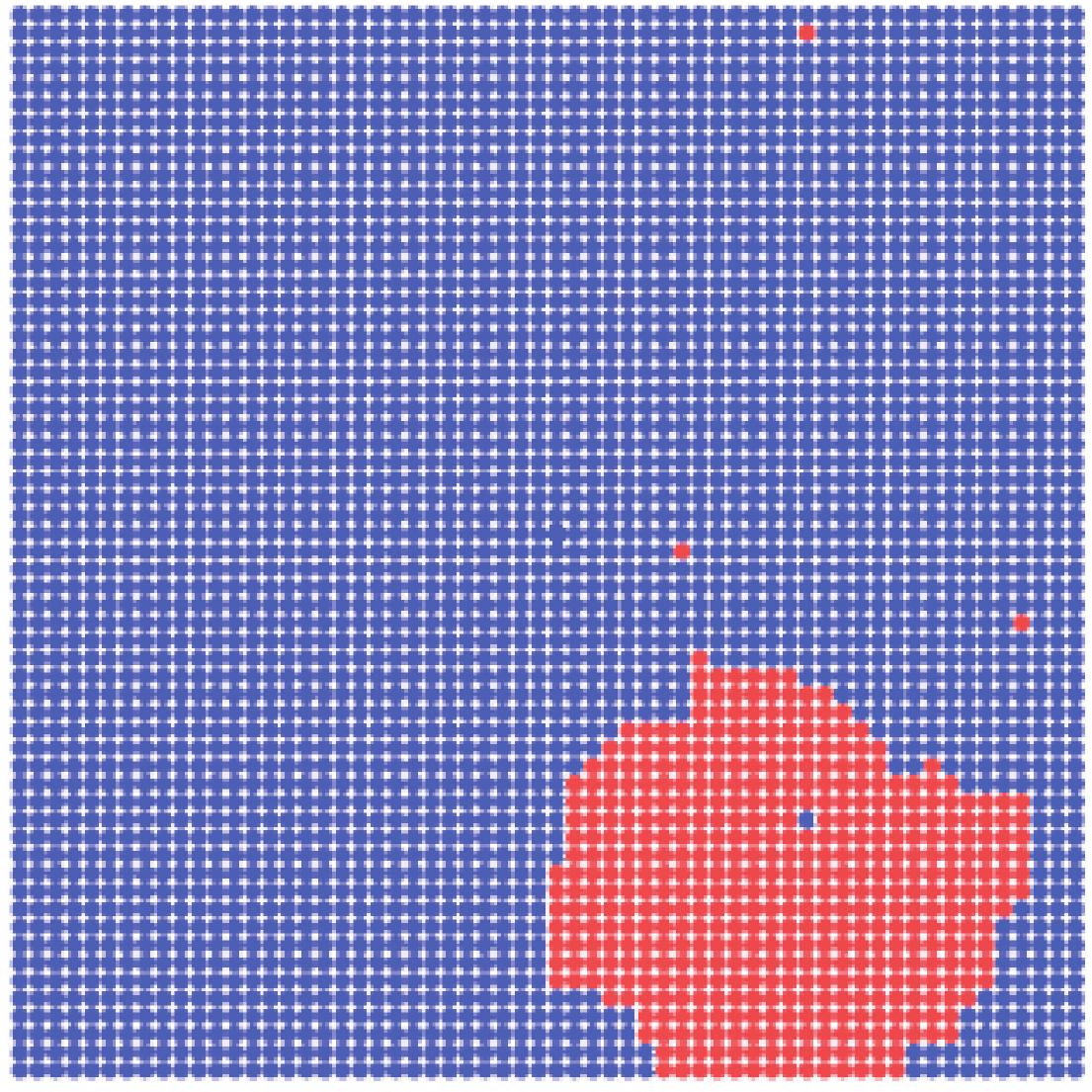}
  \caption{A typical member of TSE for the case $h_{\textnormal{S}} = -0.5J$. The lattice size is $60 \times 60$ and the surface is located at the bottom. The -1 spins are blue, and the +1 spins are red.}
  \label{fgr:4}
\end{figure}

For each slice $x$ of all sampled trajectories connecting the spin-down and spin-up states, we first generate 10 fleeting trajectories starting from $x$, and determine $p_{\uparrow}(x,t_{\textnormal{s}})$ from the fraction of paths that end in the spin-up state. Since in this approach we are dealing with the same configurations $x$ as in the above committor $p_{\uparrow (\downarrow)}(n,t_{\textnormal{s}})$ analysis, we take the same fleeting trajectory length $t_{\textnormal{s}}=1800$ time units to ensure $p_{\uparrow}(x,t_{\textnormal{s}}) + p_{\downarrow}(x,t_{\textnormal{s}}) \approx 1$.
The configuration $x$ is rejected as a member of the TSE if the calculated $p_{\uparrow}$ falls outside the $95\%$ confidence interval around $p_{\uparrow} = 0.5$. Otherwise more fleeting trajectories are generated until either $p_{\uparrow}$ falls outside this confidence interval, or an upper limit of 50 fleeting trajectories is reached. If former is the case, the configuration is rejected, otherwise it is accepted as a member of the TSE. From the trajectories connecting spin-down and spin-up states harvested with TPS, we typically found between 2000 and 3000 TSE members for each value of the surface field $h_{\textnormal{S}}$. A typical critical nucleus taken from the TSE has a rough interface and is anisotropic (see Fig.~\ref{fgr:4}). 
\begin{table}[h]
\small
  \caption{Surface fields $h_{\textnormal{S}}$ and the corresponding choices of limits for the spin-down state $n_{\downarrow}$ and spin-up state $n_{\uparrow}$.
   Also shown is a comparison between the critical cluster sizes $n^*$ obtained with TSE analysis and umbrella sampling (US).}
  \label{tbl:example}
  \begin{tabular*}{0.5\textwidth}{@{\extracolsep{\fill}} c  c  c  c  c}
    \hline
    $h_{\textnormal{S}}$ & $n_{\downarrow}$ & $n_{\uparrow}$ & TSE $\, n^{*}$ & US $\, n^{*}$\\
    \hline
    -0.1J & 20 & 600 & 288 & 285\\
    -0.2J & 20 & 600 & 324 & 325\\
    -0.3J & 20 & 600 & 357 & 355\\
    -0.4J & 20 & 700 & 395 & 405\\
    -0.5J & 20 & 700 & 430 & 415\\
    -0.6J & 20 & 700 & 454 & 445\\
    \hline
  \end{tabular*}
  \label{tbl:1}
\end{table}

For all  configurations $x$ in the TSE we can calculate the size of largest cluster $n(x)$, and hence determine its distribution. The average value of this distribution is then taken to be the size of the critical cluster $n^{*}$. Table \ref{tbl:1} shows that the TSE results are in good agreement with the corresponding values of $n^{*}$ obtained using umbrella sampling.

\subsection{Contact angles}

We use TPS to estimate the contact angle of a growing nucleus with the surface for different surface fields $h_{\textnormal{S}}$. In these simulations we set $n_{\downarrow} = 20$ and $n_{\uparrow}=1000$ and sample reactive trajectories of 1000 time units in length for surface fields $h_{\textnormal{S}} \ge -0.3 J$, and 1200 time units for surface fields $h_{\textnormal{S}} \le -0.4J$. We relax the initial trajectory with 25000 shooting moves, and then sample 2500 independent trajectories, one every 100 accepted shooting moves. In this calculation we are able to analyse more sampled trajectories than in the TSE simulations, because the latter require additional shooting of fleeting trajectories which is computationally very costly. 
\begin{figure}[h]
\centering
  \includegraphics[height=6.5cm]{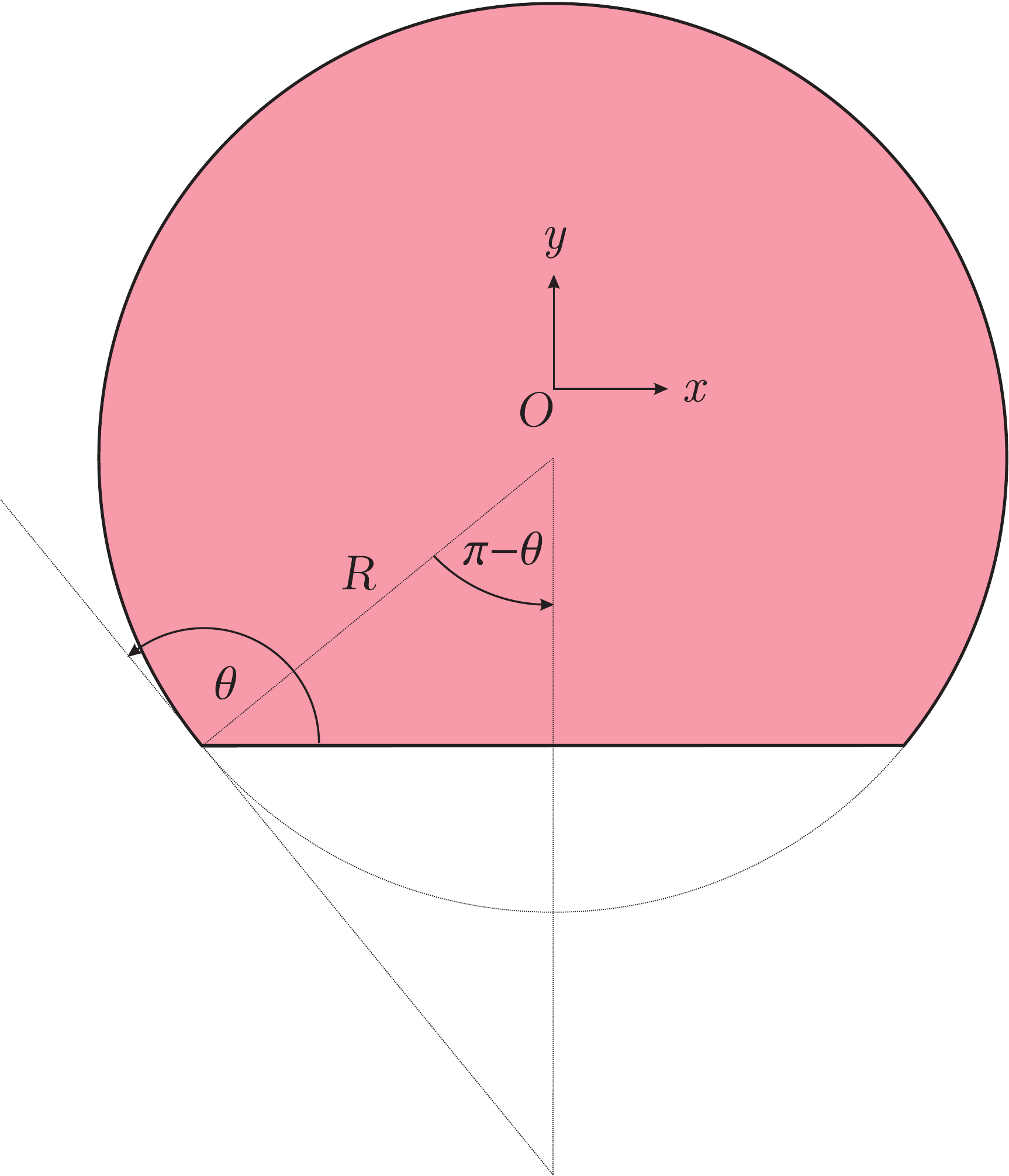}
  \caption{The shape of the circular segment can be conveniently characterised by two components of the gyration radius with respect to the coordinate system $xOy$ (with $O$ being the centre of mass). They can be calculated explicitly and expressed in terms of the angle $\theta$ and say, the radius $R$. Thus, the ratio of the squared components of the gyration radius depends only on $\theta$ as given by expression (\ref{gratio}).}
  \label{fgr:5}
\end{figure}

To determine the average contact angle $\theta_n$ of nuclei of size $n$ we proceed as follows. It is expected, on symmetry grounds, that the average cluster has the shape of a circular segment, such as that presented in Fig.~\ref{fgr:5}. 
The shape of a cluster of size $n$ can be characterised by the components $R_{\textnormal{g},x}$ and $R_{\textnormal{g},y}$ of the gyration radius, $R_{\textnormal{g}}^2 = R_{\textnormal{g},x}^2 + R_{\textnormal{g},y}^2$, with
\begin{equation}
R_{\textnormal{g},x}^2(n) = \frac{1}{n} \sum_j (x_j - x_{\textnormal{c}})^2, \quad R_{\textnormal{g},y}^2(n) = \frac{1}{n} \sum_j (y_j - y_{\textnormal{c}})^2,
\end{equation}
where $x_j$ and $y_j$ are the coordinates of $j$-th spin in the cluster $n$, and $x_{\textnormal{c}}$ and $y_{\textnormal{c}}$ are the coordinates of its centre of mass. The mean values of these quantities, $\langle R_{\textnormal{g},x}^2 \rangle$ and $\langle R_{\textnormal{g},y}^2 \rangle$, obtained by averaging over many different realisations of clusters of given size $n$, can be approximated with the squared components of the gyration radius of a homogeneous circular segment having a `mass' $n$. Since the ratio of the latter components can be calculated directly, we can write
\begin{eqnarray}
	\frac{ \langle R_{\textnormal{g},y}^2 \rangle}{\langle R_{\textnormal{g},x}^2 \rangle} \left [ 1 - \frac{2 \cos \theta_n \sin^3{\theta_n}}{3(\theta_n - \cos \theta_n \sin \theta_n)} \right ] =  \nonumber \\
	  1 - \frac{16 \sin^6{\theta_n}}{9(\theta_n - \cos \theta_n \sin \theta_n)^2} + \frac{2 \cos \theta_n \sin^3{\theta_n}}{\theta_n - \cos \theta_n \sin \theta_n},
\label{gratio}
\end{eqnarray}
where $\theta_n$ is the average contact angle of clusters of size $n$. 
This expression allows us to determine $\theta_n$ as a function of the ratio $\langle R_{\textnormal{g},y}^2 \rangle /  \langle R_{\textnormal{g},x}^2 \rangle$ measured in simulations.
\begin{figure}[h]
\centering
  \includegraphics[width=8.6cm]{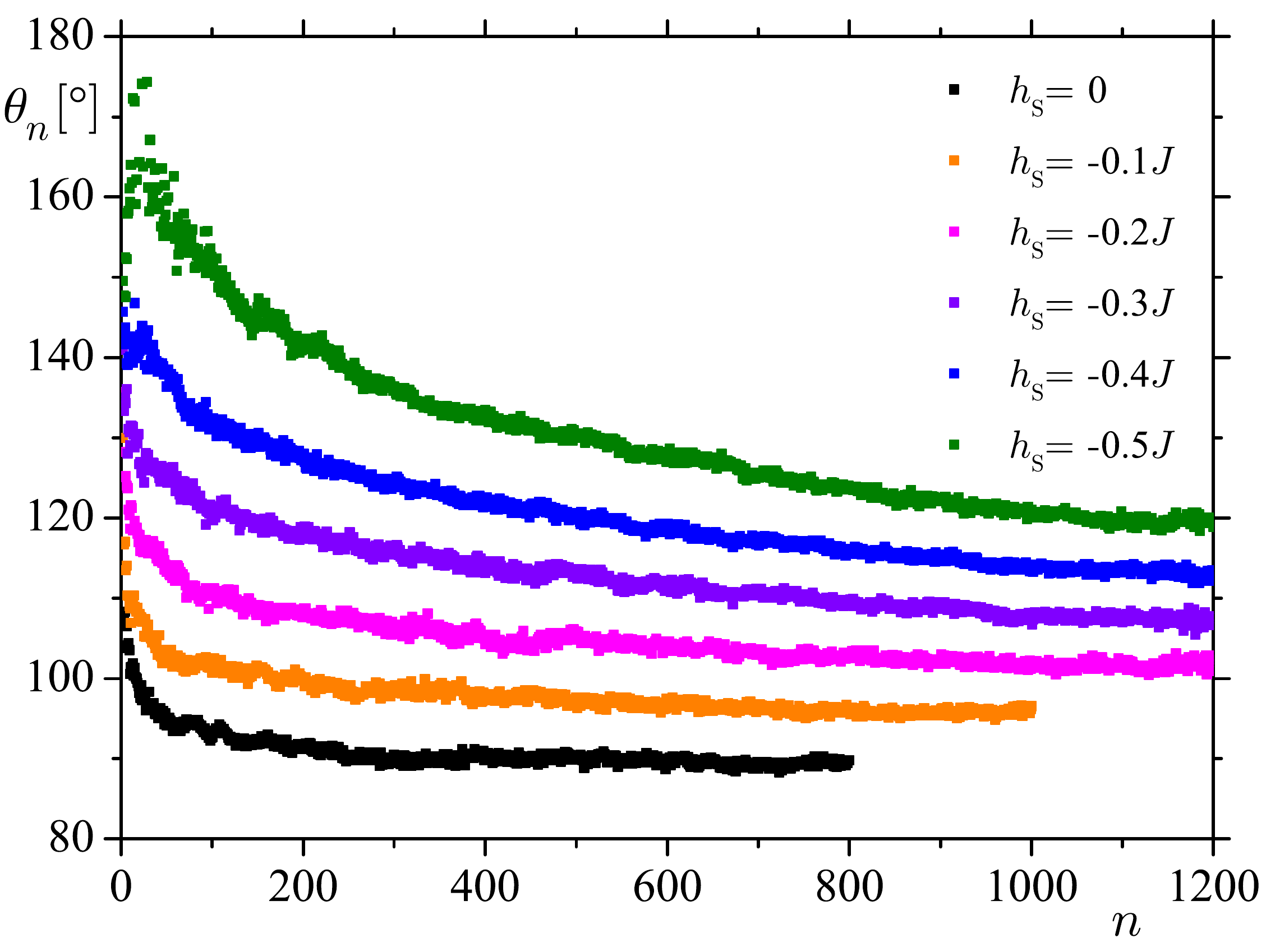}
  \caption{Average contact angle $\theta_n$ as a function of cluster size $n$ for various surface fields $h_{\textnormal{S}}$.}
  \label{fgr:6}
\end{figure}

The change of contact angle with cluster size for various surface fields $h_{\textnormal{S}}$ is shown in Fig.~\ref{fgr:6}. While for small cluster sizes the angle changes rapidly, it saturates for larger cluster sizes and converges towards a value which depends on the surface field $h_{\textnormal{S}}$.
Note that, for neutral surfaces, $h_{\textnormal{S}} = 0$, one indeed gets a contact angle of $90^{\circ}$ as predicted~\cite{Winter2009}. As the surface field decreases, the contact angle increases, reaching a value of approximately $120^{\circ}$ for $h_{\textnormal{S}} = - 0.5J$. For even smaller surface fields $h_{\textnormal{S}}$, it turns out to be difficult to measure the contact angles, because the sampled reactive trajectories connecting the spin-down and spin-up phase do not always exhibit nucleation on the surface -- a significant number of trajectories follows a homogeneous nucleation pathway in the bulk. 

\section{Nucleation from neutral pores}\label{neutral}

Before considering nucleation from hydrophobic pores, it is helpful to briefly examine the case of neutral pores, $h_{\textnormal{S}} =0$. Nucleation in neutral pores was investigated by Page and Sear~\cite{Page2006}, who found that it often occurs as a two-step process. The first of two activated processes is nucleation in the pore, followed by nucleation out of the filled pore. The authors calculated
the two nucleation rates using a forward flux sampling method~\cite{Allen2006,Allen2009}. They found that the nucleation rate in a pore increases with decreasing pore width, while the nucleation rate out of the filled pore grows with increasing pore width.

\begin{figure}[h]
\centering
  \includegraphics[width=8.6cm]{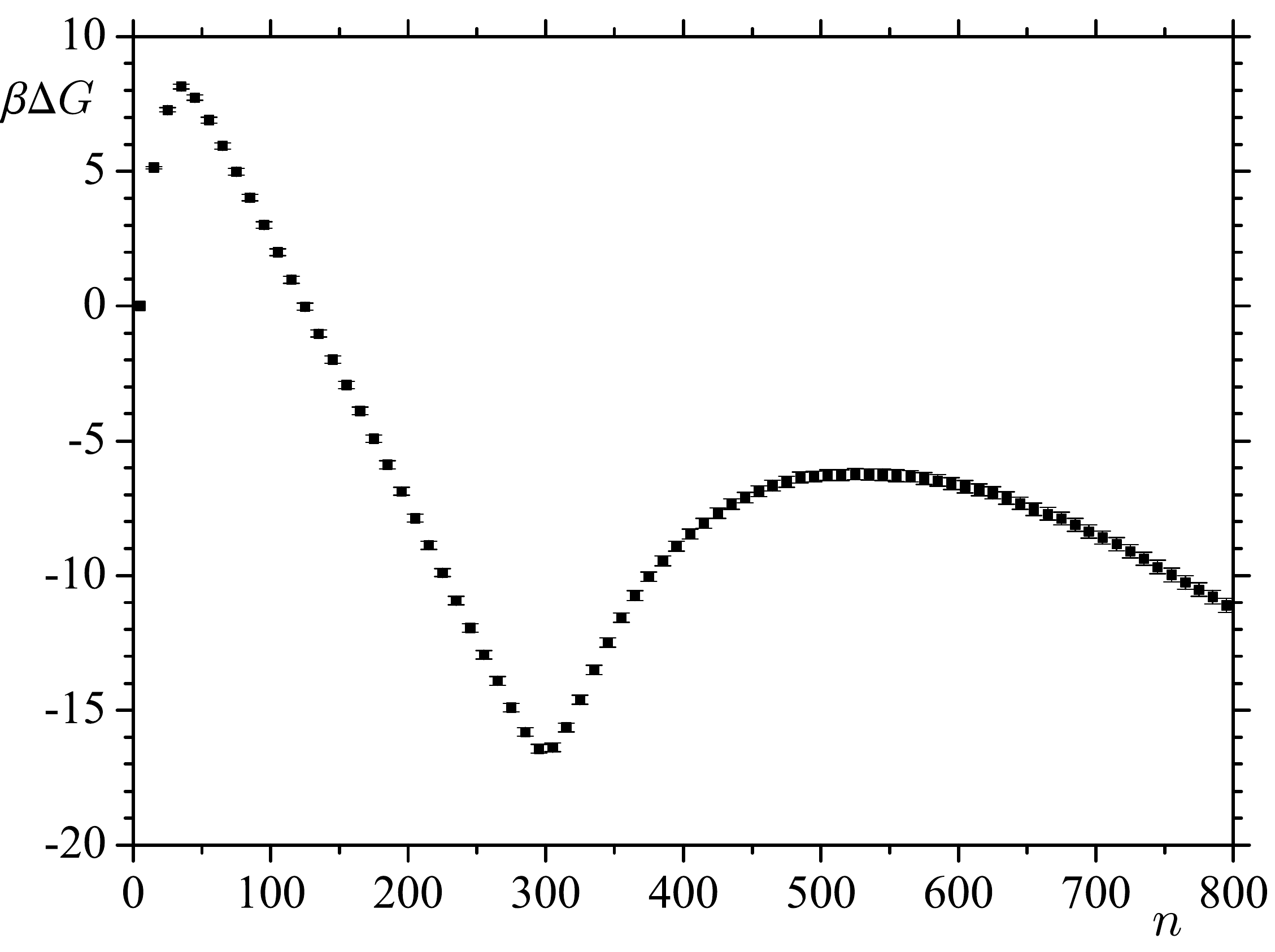}
  \caption{Free energy profile as a function of nucleus size $n$ for a neutral pore ($h_{\textnormal{S}}=0$) of size $d=30$, $w=10$.}
  \label{fgr:7}
\end{figure}

We keep the same values of $T$ and $h$ as in Section \ref{flat}, and choose the pore depth $d=30$ (Fig.~\ref{fgr:1}). 
We use umbrella sampling to calculate the free energy barriers for nucleation for various pore widths, and compare our results with the findings of Ref.  [\onlinecite{Page2006}]. An example of the dependence of the free energy profile on the cluster size $n$ for two-step nucleation, obtained for $w=10$, is shown in Fig.~\ref{fgr:7}. The nucleation always starts in one of the bottom corners of the pore. From a corner a small nucleus grows until it reaches some critical size, which corresponds to the left maximum of $\Delta G$ in Fig.~\ref{fgr:7}. The nucleus then proceeds to grow without any free energy cost until the whole pore is filled with +1 spins. Note that there is a minimum in the free energy at $n \approx 300$ which is equal to $d \times w$. Once the pore is full, there is a second barrier against nucleation of the spin-up phase outside the pore indicated by the right maximum of $\Delta G$ in Fig.~\ref{fgr:7}.

\begin{figure}[h]
\centering
  \includegraphics[width=8.6cm]{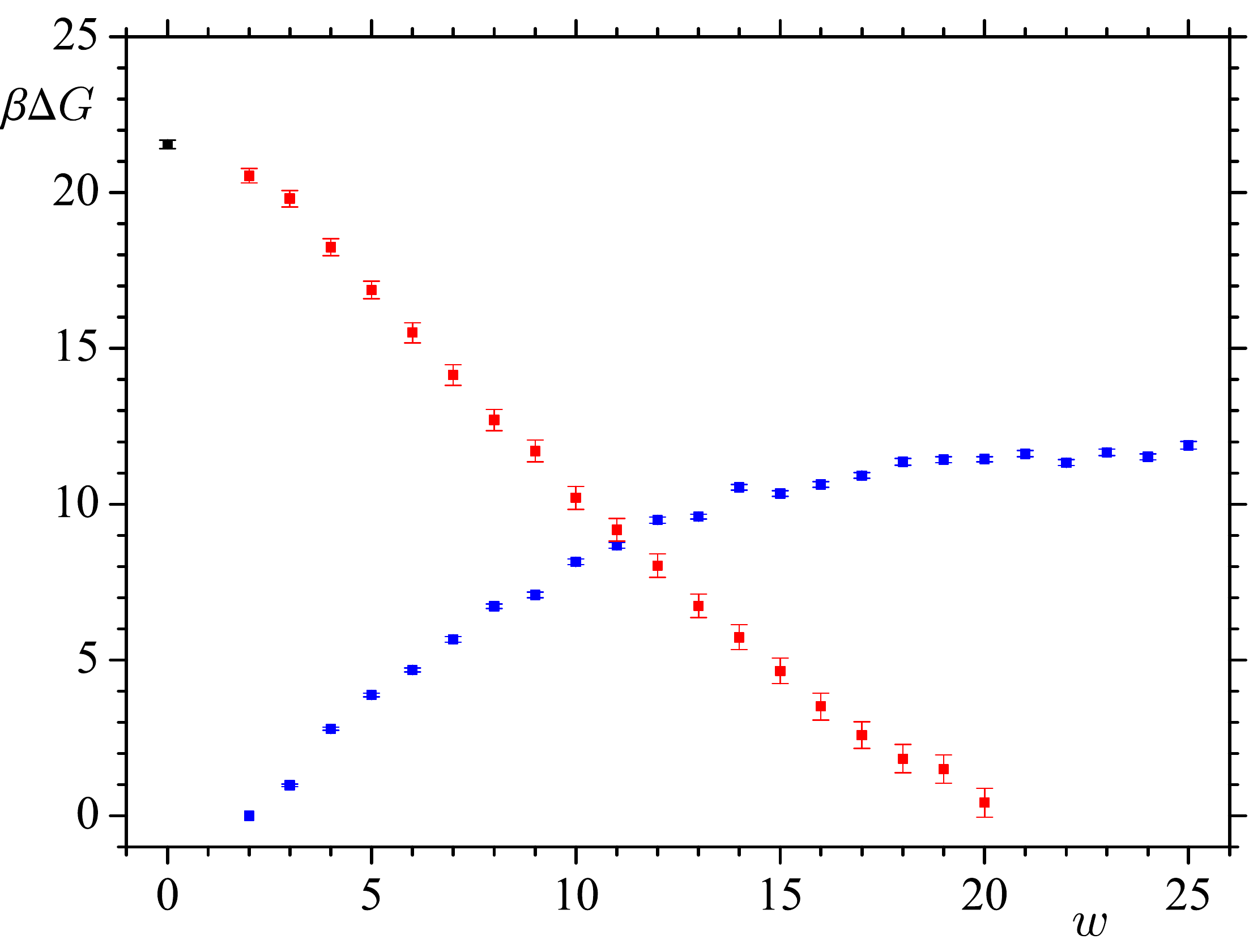}
  \caption{Nucleation barrier heights for nucleation in (blue) and out (red) of the neutral pores as a function of $w$; pore depth is $d=30$. The nucleation barrier height for a neutral flat surface is marked with a black point.}
  \label{fgr:8}
\end{figure}

However, the nucleation need not always occur via two steps. Nucleation barrier heights for nucleation in the pore and out of the filled pore for a range of pore widths $w$ are shown in Fig.~\ref{fgr:8}. The blue curve shows that the nucleation barrier in a pore is negligibly small for small pore widths, but grows and eventually saturates with the increase of $w$. This is because the nucleation always starts in one of the bottom corners and, for small pore widths, the nucleus feels the presence of the other corner, which speeds up the nucleation, whereas for wider pores the corners act independently.
This result is in a good qualitative agreement with the findings of Ref. [\onlinecite{Page2006}], because the increase of barrier heights manifests itself as a decrease of nucleation rates in the pore.

Looking at the barriers for nucleation outside a filled pore (red curve in Fig.~\ref{fgr:8}), one observes that for small $w$ the barrier height approaches that for nucleation on a flat surface. With the increase of pore width the barrier heights decrease until they eventually become negligibly small. Again, this agrees well with the results of Page and Sear~\cite{Page2006}, who found that the rates for nucleation out of a filled pore grow with $w$.
In addition, these authors found that there is a maximum in the overall nucleation rate from the pore for $w=12$, which is the point where the nucleation rates in the pore and out of the filled pore are approximately equal. In our umbrella sampling simulations the two barriers are found to be the same at $w=11$. The small discrepancy is most likely due to other (subdominant) factors affecting the nucleation rates~\cite{Debenedetti1996}. 

\section{Nucleation from hydrophobic pores}\label{hydrophobic}

Having verified that our method works well for neutral pores, we move to the case of hydrophobic pores, which have walls with surface fields $h_{\textnormal{S}}<0$. In what follows, we keep the same values of $T$, $h$ and $d$, and choose $h_{\textnormal{S}} = -0.5J$, which corresponds to nuclei with a contact angle of $\theta \approx 120^{\circ}$ on a flat surface (see Fig.~\ref{fgr:6}).
We apply umbrella sampling to calculate free energy barriers for nucleation for various pore widths $w$. We find that there are three different dynamical pathways to nucleation, which are realised for three different $w$ intervals. 

\begin{figure}[h]
\centering
  \includegraphics[width=8.6cm]{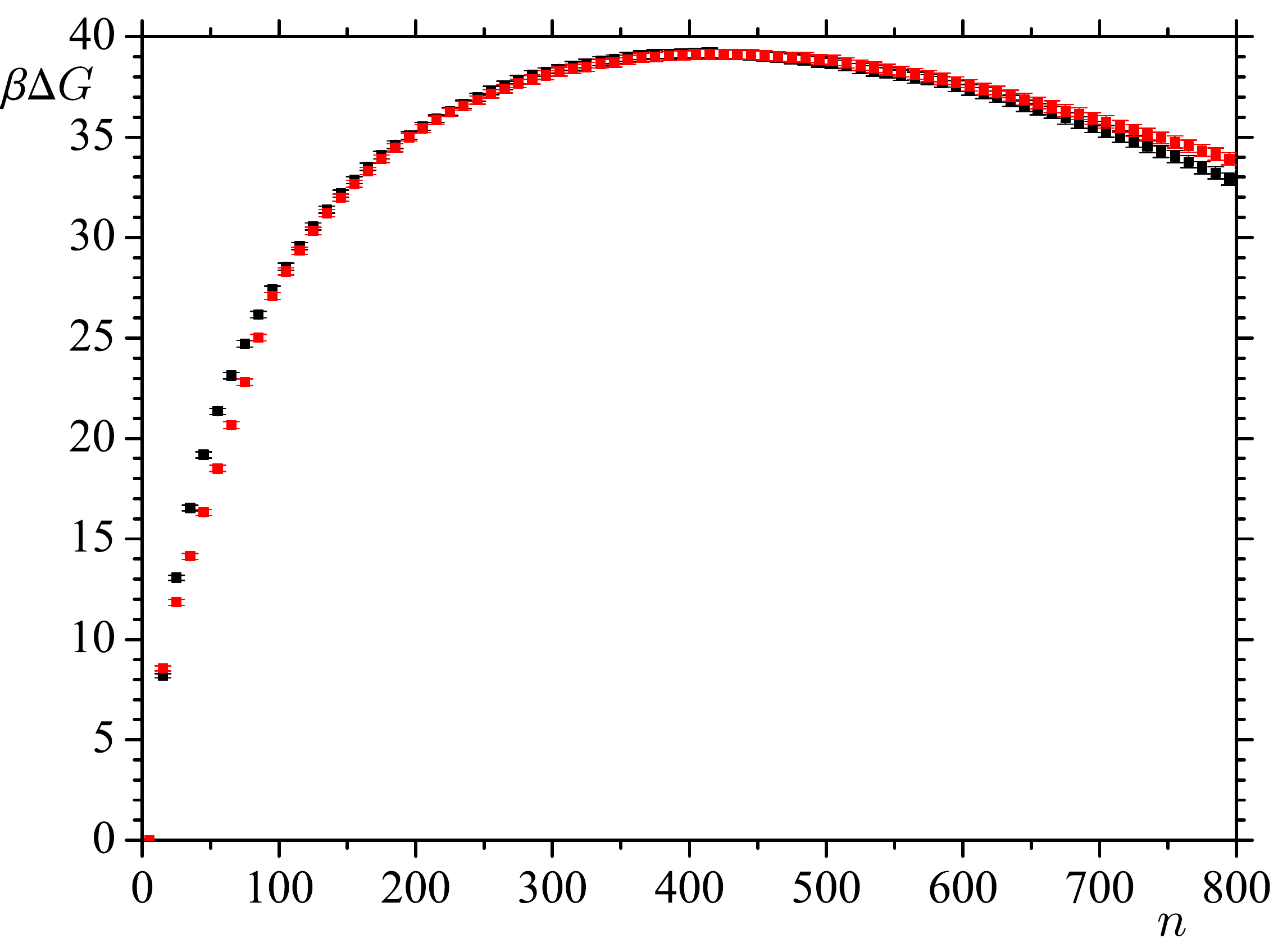}
  \caption{Free energy barrier for nucleation on a flat surface (black), and in a pore of size $d=30$, $w=5$ (red). In both cases $h_{\textnormal{S}}= -0.5J$.}
  \label{fgr:9}
\end{figure}

\begin{figure}[h]
\centering
  \includegraphics[width=7.4cm]{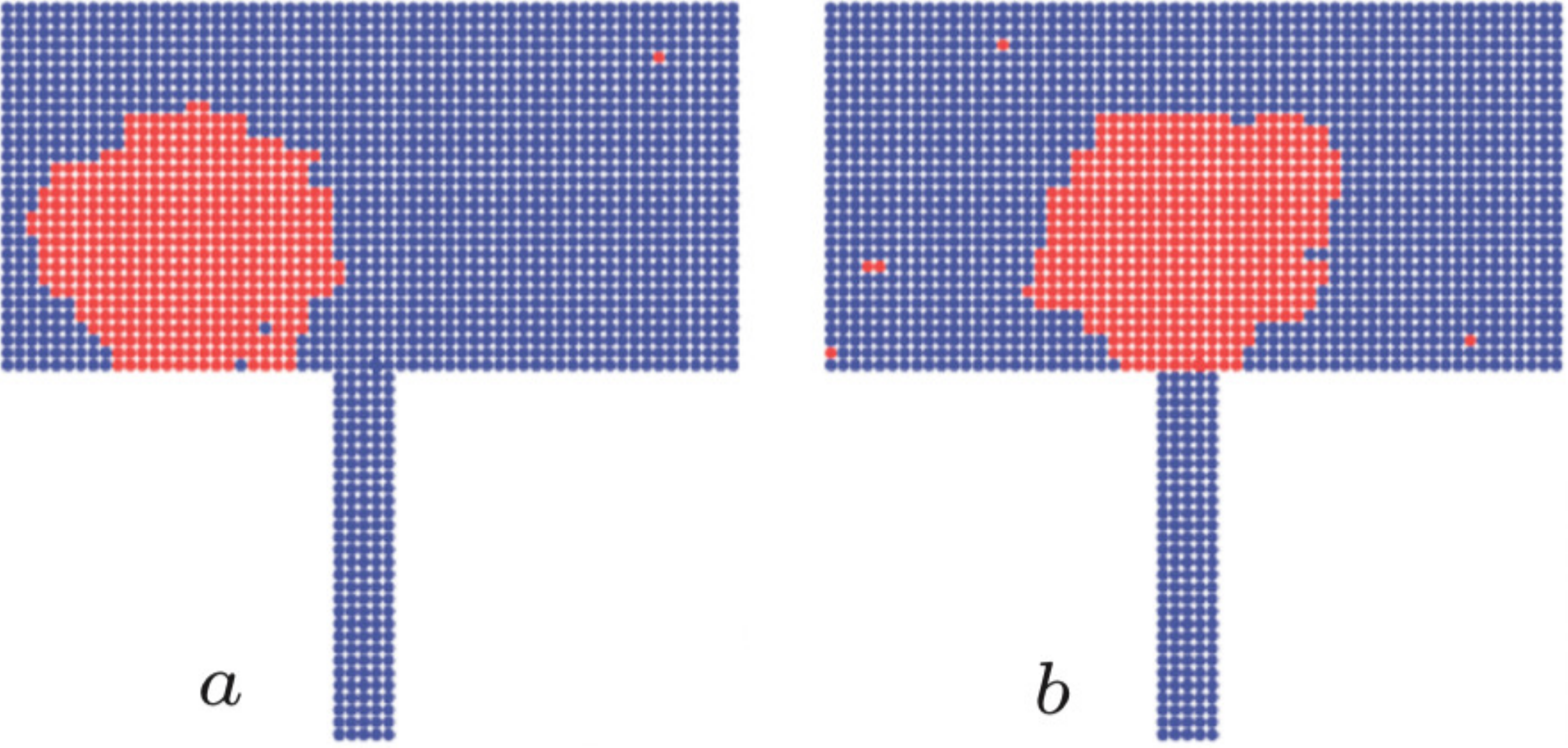}
  \caption{Typical examples of critical nuclei for the pore of size $d=30$, $w=5$: (a) nucleus on the surface outside the pore, (b) nucleus spanning the pore.}
  \label{fgr:10}
\end{figure}

The simplest nucleation pathway is found for $w \le 5$. In this case small pre-critical nuclei grow and shrink both in the pore and on the surface outside the pore, but eventually the critical nucleus is formed on the surface outside the empty pore.
This nucleation pathway has already been observed~\cite{Hedges2012} in simulations of 3D pores.
As expected, the free energy barrier for nucleation looks very similar to that on a flat surface: A comparison for $w=5$ is shown in Fig.~\ref{fgr:9}.
Although the vast majority of critical nuclei are formed on the surface outside the pore (Fig.~\ref{fgr:10}a), one occasionally finds a nucleus spanning the pore (Fig.~\ref{fgr:10}b). 

\begin{figure}[h]
\centering
  \includegraphics[width=8.6cm]{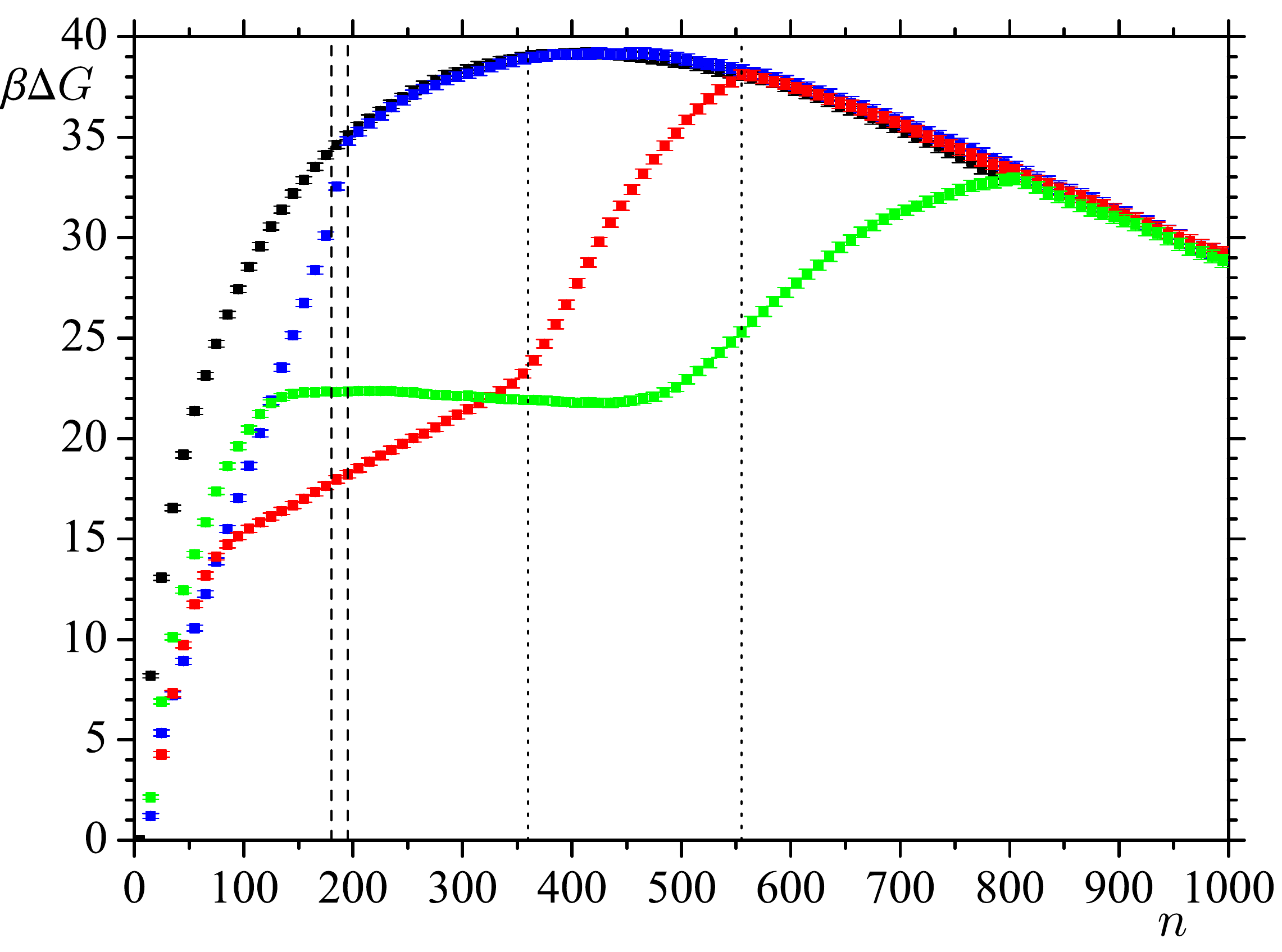}
  \caption{Free energy barriers for nucleation from hydrophobic pores of depth $d=30$ and widths $w=6$ (blue), $w=12$ (red) and $w=16$ (green). The free energy barrier for nucleation on a flat surface is shown in black.} 
  \label{fgr:11}
\end{figure}

The most interesting behaviour emerges in the range of pore widths $6 \le w \le 16$. Typical free energy barriers are shown in Fig.~\ref{fgr:11}. The nucleation mechanism for $w=6$ (blue curve) significantly differs from that encountered for widths $w \le 5$. A pre-critical nucleus starts to appear in one of the bottom corners of the pore. The nucleus continues to grow and eventually the whole pore is filled with +1 spins. However, on this growth pathway the nucleus does not attain a critical value, which manifests itself as the absence of a maximum in the free energy in the region of cluster sizes between $n=0$ and $n= 30 \times 6 = 180$. The full pore state, $n= 180$ corresponds to the left dashed vertical line in Fig.~\ref{fgr:11}. At this point there is a noticeable change in the slope of the free energy curve. The nucleus then starts to grow outside the filled pore and does so until it reaches $n \approx195$ (marked with the right dashed vertical line). Inspection of the cluster configurations shows that at this point the nucleus is expelled from the pore and ends up spanning the pore (Fig.~\ref{fgr:10}b); during the ejection the cluster size remains approximately unchanged. Thus the pre-critical nucleus exhibits a pore emptying transition. The pre-critical nucleus then continues to grow outside the pore, following the free energy curve for nucleation on a flat surface, and reaching the maximum of the free energy curve corresponding to its critical size at $n \approx450$. This dynamical pathway is qualitatively similar for all pore widths $6 \le w \le 11$. 

\begin{figure}[h]
\centering
  \includegraphics[width=7.4cm]{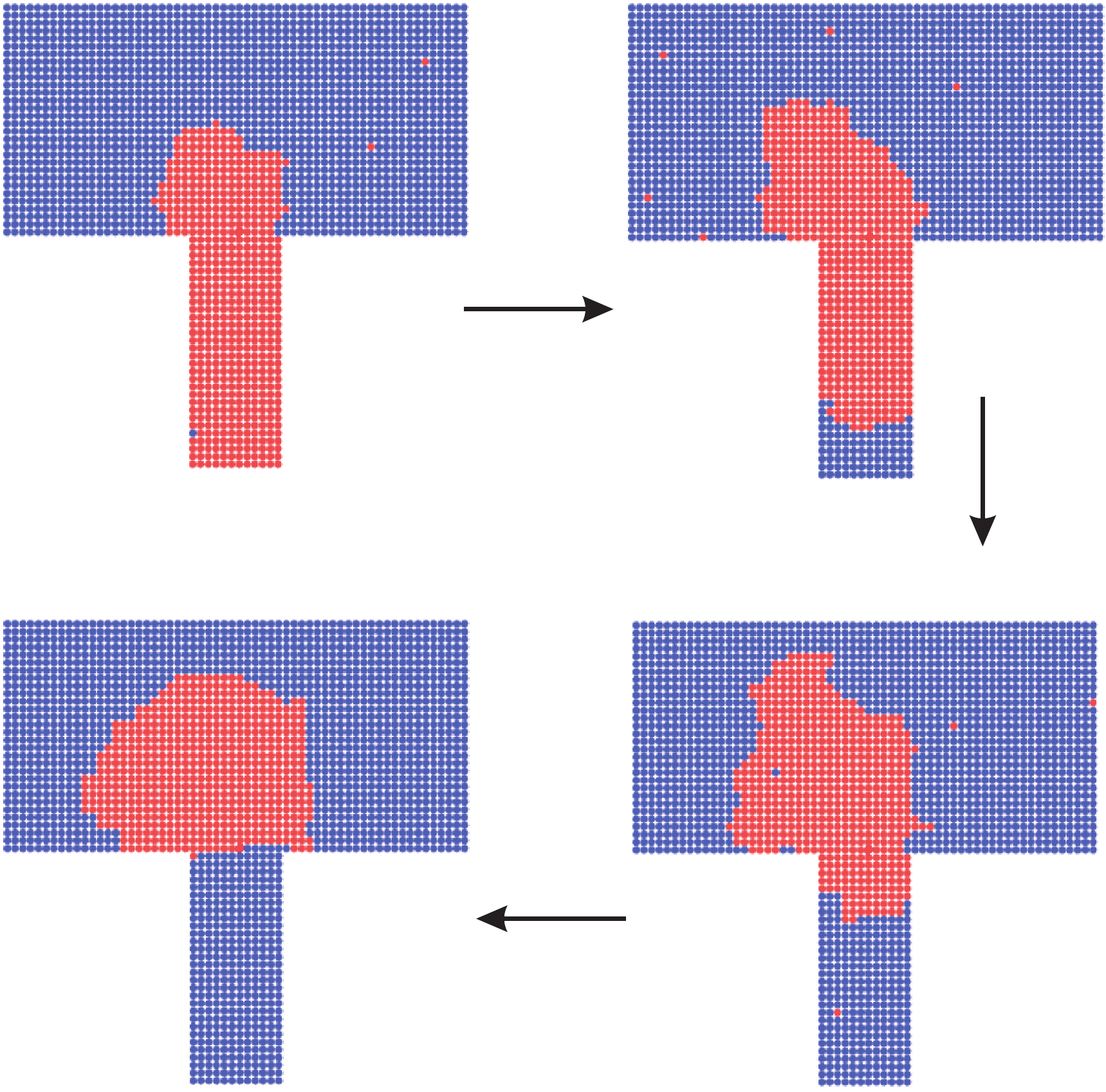}
  \caption{Snapshots of the pore emptying transition for the pore of size $d=30$, $w=12$. In all snapshots the size of the nucleus is $n^* \approx 555$.}
  \label{fgr:12}
\end{figure}

For $w=12$ (red curve in Fig.~\ref{fgr:11}) there is a small change in behaviour. The full pore state, corresponding to a change in the rate of increase of the free energy with $n$, is marked by the left dotted vertical line at $n = 360$. For this pore width the free energy maximum $n \approx 555$ (marked by the right dotted vertical line) coincides with the point where the nucleus is expelled from the pore; typical cluster configurations during the ejection are shown in Fig.~\ref{fgr:12}. Thus the nucleus attains its critical size immediately after expulsion, and then continues to grow following the free energy barrier for nucleation on a flat surface (black curve in Fig.~\ref{fgr:11}). 
A qualitatively similar dynamical pathway is also found for larger pore widths up to $w=16$.

\begin{figure}[h]
\centering
  \includegraphics[width=8.6cm]{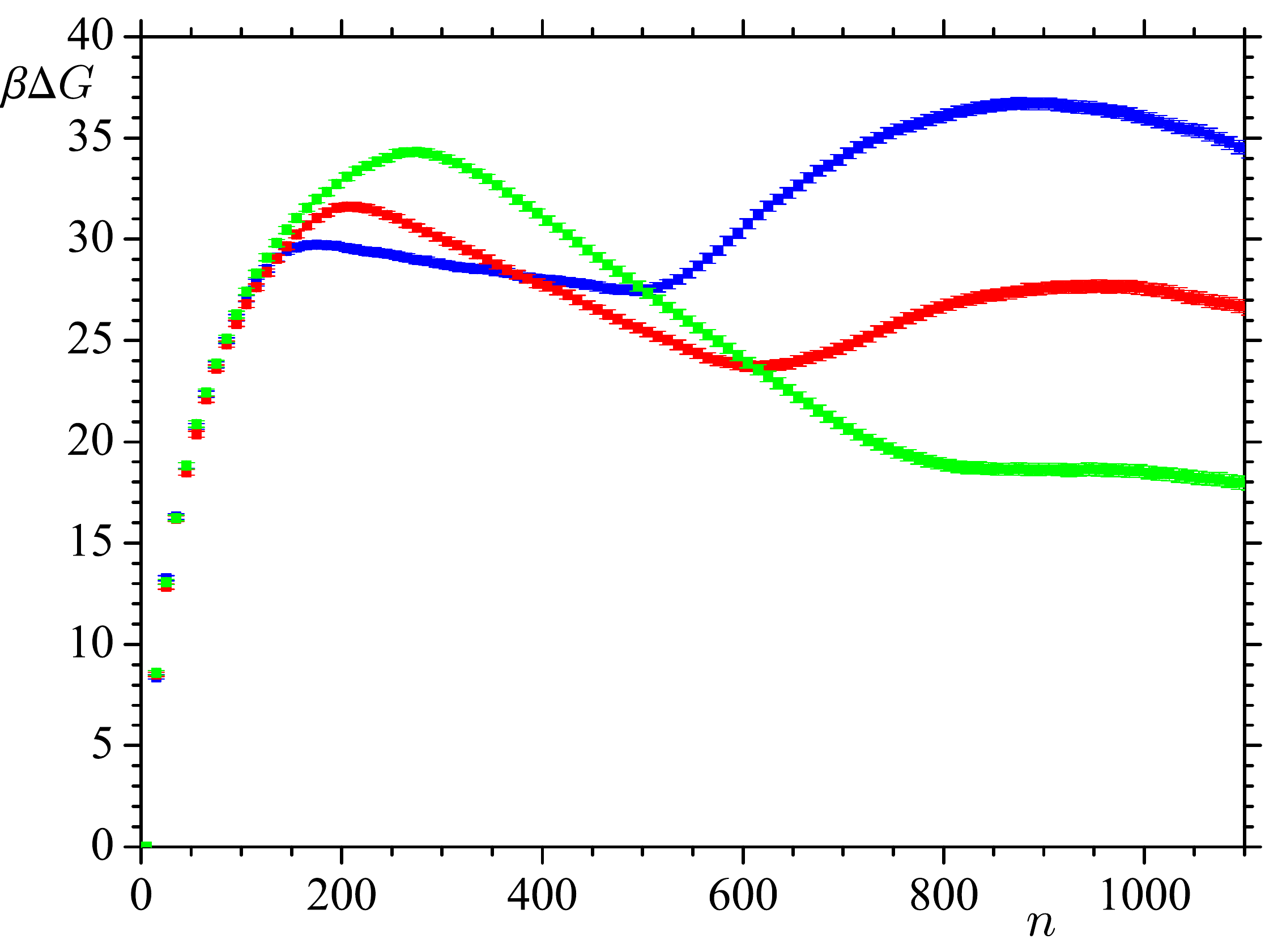}
  \caption{Free energy barriers for nucleation from hydrophobic pores of depth $d=30$ and widths: $w=17$ (blue), $w=20$ (red) and $w=24$ (green).}
  \label{fgr:13}
\end{figure}

\begin{figure}[h]
\centering
  \includegraphics[width=7.4cm]{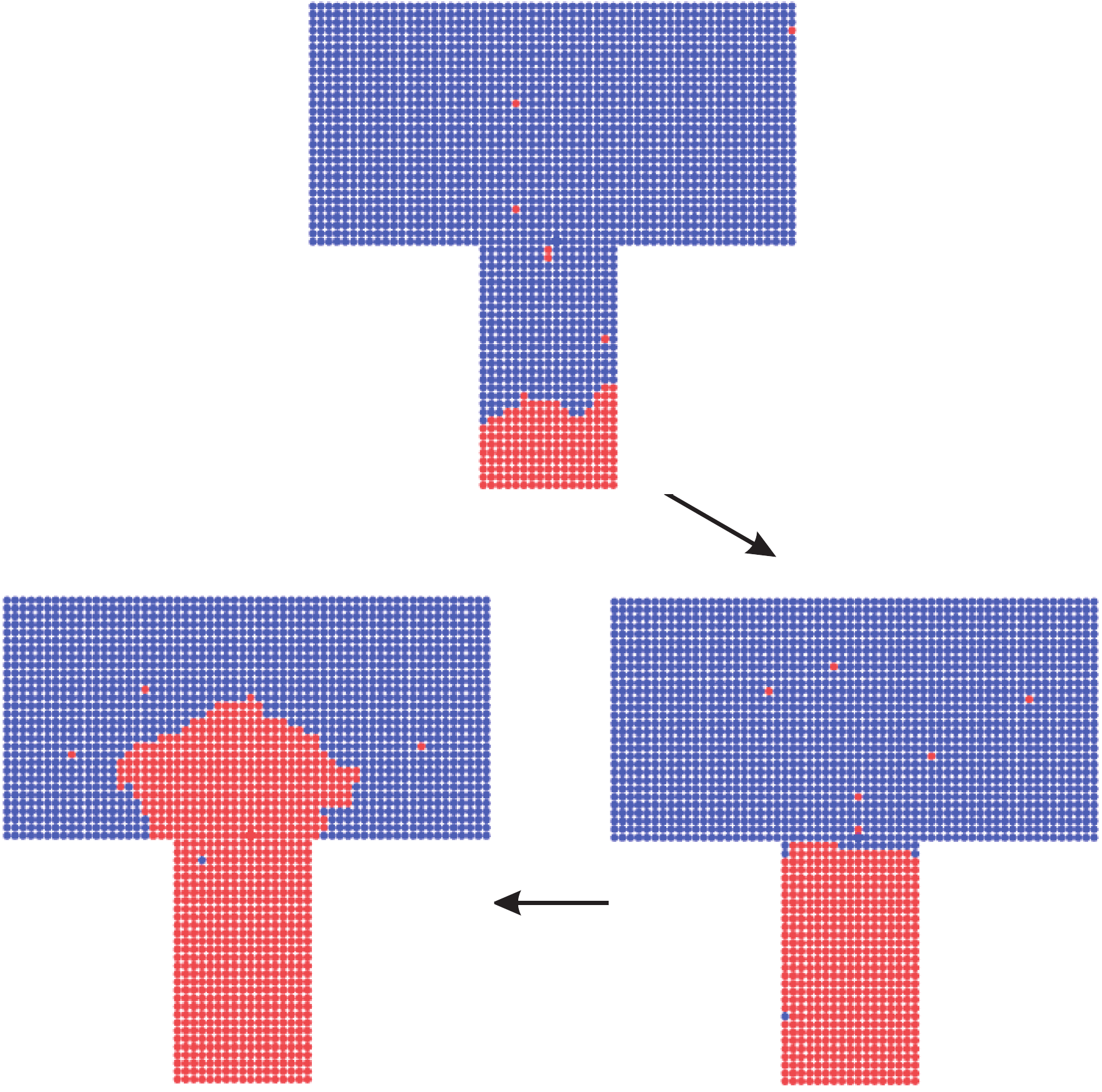}
  \caption{A two-step nucleation process for the pore of size $d=30$, $w=17$.}
  \label{fgr:14}
\end{figure}

This regime ends at $w=16$ (green curve in Fig.~\ref{fgr:11}). For this pore width the point on free energy curve corresponding to the full pore, $n \approx 480$, becomes a minimum, while at the same time a maximum appears for smaller cluster sizes.
However, the free energy difference between the maximum and minimum is negligibly small (smaller than the typical error bars), and the nucleus again empties the pore during nucleation. 
For $w=17$ the minimum becomes more pronounced (blue curve in Fig.~\ref{fgr:13}) and the system crosses over to a new regime, where it follows a two-step nucleation pathway. As before pre-critical nuclei appear in one of the bottom corners of the pore, and then continue to grow. However the nuclei now achieve a critical size for a value of $n$ smaller than the pore size. This critical size corresponds to the left maximum of the blue curve in Fig.~\ref{fgr:13}. After attaining the critical size, the nuclei grow further without any free energy cost until they fill the pore. The minimum in the free energy curves is located at approximately $n= 30 \times 17 = 510$, the pore size. Finally, there is a second nucleation barrier for growth outside the filled pore, which corresponds to the right maximum in Fig. \ref{fgr:13} (blue curve).  Snapshots of the growing cluster for $w=17$ are shown in Fig.~\ref{fgr:14}.
This nucleation process is reminiscent of the two-step nucleation from neutral pores analysed in Section \ref{neutral}.

The barrier to nucleation within the pore increases with the pore width $w$ and eventually saturates, while the barrier to nucleation outside the filled pore decreases with $w$. The two-step nucleation is observed for pore widths $17 \le w \le 23$. At $w=24$ (shown in green in Fig.~\ref{fgr:13}) the right maximum disappears and the nucleation occurs in one step. There is a critical nucleus smaller than the pore size, after which the cluster continues to grow, both in and out of pore, without any additional free energy cost.

\section{Summary}

To summarise, we have used umbrella sampling and transition path sampling Monte Carlo simulations performed on a simple model system, the 2D Ising model, to provide generic picture of nucleation in hydrophobic nanopores.
We find three regimes. For narrow pores, the nucleation occurs on the flat surface, almost unaffected by the presence of the pores. For wide pores the pores fill and then act as a seed for nucleation of a critical cluster on the surface.
For pores of intermediate widths nucleation commences in the pores but, as the cluster grows across the surface, there is a sudden pore emptying transition before the nucleus has reached its critical size. 

Our results suggest that the nanoscale tomography of a surface will be key in controlling its water-repellency. In particular, narrower pores will more readily empty during nucleation to give the Cassie-Baxter configuration, associated with
fast droplet shedding, and relevant to the enhanced performance of condensers, anti-icing substrates and self-cleaning materials. Extensions to 3D and to more realistic liquid models are needed to help to quantify these effects, but will be
extremely demanding of computational resources. 

\section*{Acknowledgments}

We are grateful to Richard Sear for helpful discussions. This work was supported through funding from the ERC Advanced Grant 291234 MiCE.

\end{document}